\documentclass[a4paper]{article}

\usepackage{icrc2013}

\title{VERITAS Long-Term Observations of Hard Spectrum Blazars}

\shorttitle{Hard Spectrum Blazars}

\authors{
Arun S. Madhavan$^{1}$,
for the VERITAS Collaboration.
}

\afiliations{
$^1$ Department of Physics \& Astronomy, Iowa State University, Ames, IA, 50014, USA \\
\scriptsize{
}
}

\email{madhavan@iastate.edu}

\abstract{The VERITAS collaboration has approved long-term observations on several distant, hard-spectrum blazars.  We present first results from VERITAS long-term observations of 1ES1218+304, 1ES0229+200, and 1ES0414+009.  Gamma-ray observations of distant, hard-spectrum blazars has emerged as an effective, indirect probe of the universe's diffuse extragalactic background light (EBL), due to the extinction of gamma-rays via the pair production interaction $\gamma_{TeV}\gamma_{EBL}\rightarrow e^{+}e^{-}$, which is expected to produce absorption features in distant gamma-ray sources.  The sources presented here emit spectra that are well fit by power laws.}


\begin{document}
\maketitle

\section{Introduction}

Our study of VHE emission from hard-spectrum blazars is motivated by a technique for probing the universe's extragalactic background light via extinction of gamma-rays from distant AGN \cite{bib:GouldSchreder}.  The extragalactic background light (EBL) is a diffuse cosmological radiation field, second in intensity only to the cosmic microwave background.  The EBL consists of all light emitted since the epoch of recombination and includes contributions from all nuclear and gravitational processes throughout the universe \cite{bib:KrennrichOrrEBLPaper}, including star formation, accretion of matter onto supermassive black holes, exotic energy releases, and light from population III stars.  The spectrum of the EBL has also been affected by cosmic expansion and absorption/re-radiation by dust and polycyclic aromatic hydrocarbons.  The EBL thus encodes important information about the history of the early universe after recombination, and its measurement is of great importance in the field of cosmology.  Although direct observations of the EBL are hindered by the infrared foreground contributed by the zodiacal light, it is possible to indirectly probe the EBL through observations of extragalactic gamma-rays.  Recent analyses of gamma-ray data from blazars has resulted in the calculation of upper limits on the EBL intensity by \cite{bib:MayerEBLPaper}.  \cite{bib:HessEBL} and \cite{bib:AckermannEBLPaper} have also used gamma-ray data from AGN to measure the spectrum of the EBL.

\subsection{Motivation for Deep Blazar Observations}

Gamma-rays from active galactic nuclei (AGN) interact with EBL photons via  $\gamma_{TeV}\gamma_{EBL}\rightarrow e^{+}e^{-}$.  The cross-section for this interaction is energy dependent.  Specifically, the $\gamma\gamma$ pair production cross-section peaks at an EBL photon wavelength of $\lambda_{max}/\mu m = 1.24(E_{\gamma}/TeV)$.  It is thus possible to probe the spectrum of the EBL by searching for absorption features in blazar spectra.  Since the EBL exhibits a trough in the mid-IR, the optical depth for EBL absorption is expected to flatten at $\sim$1TeV, which can lead to a break in the observed VHE spectrum of an AGN (\cite{bib:OrrEtAl}).  Because hard-spectrum blazars produce emission at multi-TeV energies, it is expected that spectral measurements can be taken sufficiently far above and below the 1 TeV point to detect a spectral break.  Gamma-rays at sub-TeV energies are extincted primarily by EBL photons in the optical and near-IR regime, whereas multi-TeV gamma rays are extincted by EBL photons in the mid- to far-IR regime. Under the assumption that observed hard-spectrum blazars emit an intrinsic power-law spectrum, the extent of an observed spectral break is dependent on the ratio of the near- to mid-IR power of the EBL.

The energy-dependent optical depth of the EBL is a function of redshift; therefore, it is necessary to search for spectral breaks in blazars over a range of redshifts.  A multi-blazar study is also advantageous in that while the spectral break measured from a single source may not be statistically significant, several blazars can be used to deduce a redshift-dependent spectral break and arrive at a more significant result.  Our choice of \textit{distant} blazars in particular is such that all gamma-ray signals are more likely to traverse a sufficient distance to produce detectable absorption features.

\section{Blazar Observations with VERITAS}

VERITAS is an array of four ground-based gamma-ray telescopes located at The Fred Lawrence Whipple Observatory in Arizona.  The array employs the imaging atmospheric Cherenkov technique, and is sensitive to gamma-rays in the energy range of 85 GeV to 30 TeV.  The telescope reflectors are modeled after the Davies-Cotton design and include 350 hexagonal mirror facets.  Each camera is equipped with 499 photomultiplier tubes with data acquisition performed by 500 MHz FADCs.  Telescope 1 was relocated in the Summer of 2009 to increase the sensitivity of the array.  The  level 2 telescope trigger was upgraded in November of 2011, followed by an upgrade of all camera PMTs in the summer of 2012.  The instrument's energy resolution is 15\% to 25\% for observations analyzed with standard techniques.

The VERITAS collaboration has approved the sources presented here for long-term observations.  Because these sources are expected to exhibit detectable fluxes at multi-TeV energies, they are appropriate candidates for probing the optically thick regime of the EBL.  We present analysis of 1ES1218+304, 1ES0229+200, and 1ES0414+009.  Available observations have been included from all four epochs of the detector (before and after the telescope 1 move, after the trigger upgrade, and after the camera upgrade).  The blazars were observed with the source displaced $0.5^{\circ}$ from the camera center so as to obtain a background estimate that does not depend on the instrument's radially-dependent acceptance function.  Table \ref{tab:SourceInfo} shows the total number of hours taken on each source under moonless or low moonlight nights, the hours remaining after quality cuts, and the strength of the source in units of the Crab Nebula flux.

\begin{table}[h]
\begin{center}
\begin{tabular}{|p{2cm}|p{1.7cm}|p{1.5cm}|p{1cm}|}
\hline Source & Observation Hours & After Quality Cuts & Flux (C.U.) \\ \hline
1ES1218+304   & 107  & 85  & 8\%\\ \hline
1ES0229+200   & 79  & 54 & 2\%\\ \hline
1ES0414+009 & 84 & 56 & 2\%\\ \hline
\end{tabular}
\caption{Source Information.  Observation times indicate data taken under low light conditions using standard settings.}
\label{tab:SourceInfo}
\end{center}
\end{table}

\section{Analysis and Discussion}

Data have been analyzed with VEGAS, one of the standard VERITAS analysis packages.  All data runs are corrected for differences in PMT gain and signal cable length by applying parameters extracted from nightly calibration runs.  Shower images from individual telescopes are cleaned and then parameterized using the method described by Hillas \cite{bib:HillasParameters}.  Images from all participating telescopes are combined to reconstruct shower directions on the sky and impact locations on the ground.

A database of Monte Carlo gamma-ray simulations is consulted to reconstruct the energy of each shower from the image \textit{size} in each telescope, correcting for impact location and averaging over telescopes.  Image \textit{length} and \textit{width} from each telescope are compared with simulated gamma-rays observed at comparable zenith angles, azimuths, and night sky noise levels, and are used to establish selection cuts to suppress the hadronic background.

Finally, a reflected regions analysis (as described in \cite{bib:WobbleAnalysis}) is performed to subtract the remaining background and extract the gamma-ray excess.  The wobble analysis is performed with an ON region of size $\theta=0.1^{\circ}$.  The background estimate is calculated from six OFF regions of equal size to the ON region.  The gamma-ray excess is binned in energy, and the effective collection area (derived from simulations) is unfolded in order to derive the energy spectrum.

The analysis results from each source are summarized in table \ref{tab:SummaryOfAnalysis}.

 \subsection{Run selection}
 
Prior to analyzing each blazar, an initial source runlist was compiled, consisting of all runs not flagged as unusable.  Runs were then removed based on weather, hardware performance, and observers' reports.  Runs taken under moonlight conditions using the camera UV filters or reduced high voltage were also removed from the analysis.  Run times with temporarily reduced rates (due to weather) were removed as well.
 
 \subsection{Analysis of 1ES1218+304}
 
\begin{figure}[t]
  \centering
  \includegraphics[width=0.45\textwidth]{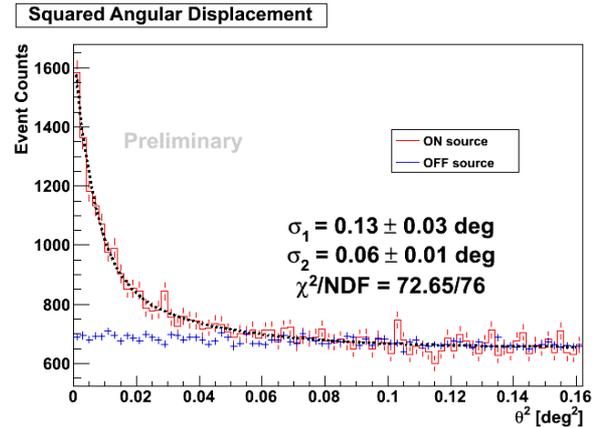}
  \caption{Squared angular displacement of events for 1ES1218+304.  The distribution of ON counts is a good fit to the PSF, indicating that 1ES1218+304 is a point source.}
  \label{pic:1ES1218Theta2}
 \end{figure}

The blazar 1ES1218+304 has been observed by VERITAS from December of 2008 until the current 2012-2013 observing season.  1ES1218+304 is located at a redshift of $z=0.182$ and is a relatively bright source with observed short-timescale variability \cite{bib:1218Asif}.  86 hours out of 107 available hours of data remain after run selection and quality cuts.  A total of 2888 excess gamma-ray events were detected, with a cumulative statistical significance of 40 standard deviations above the background.  The distribution of events in squared angular displacement from the putative source location is shown in figure \ref{pic:1ES1218Theta2}.  A double exponential model of the PSF\footnote{$f(\theta) = Ae^{-\frac{\theta^{2}}{2\sigma_{1}^{2}}} + Be^{-\frac{\theta^{2}}{2\sigma_{2}^{2}}}+C$} has been fit to the ON source counts.  This model provides a good fit, thus demonstrating that 1ES1218+304 is indeed a point source.
 
\begin{figure}[t]
  \centering
  \includegraphics[width=0.5\textwidth]{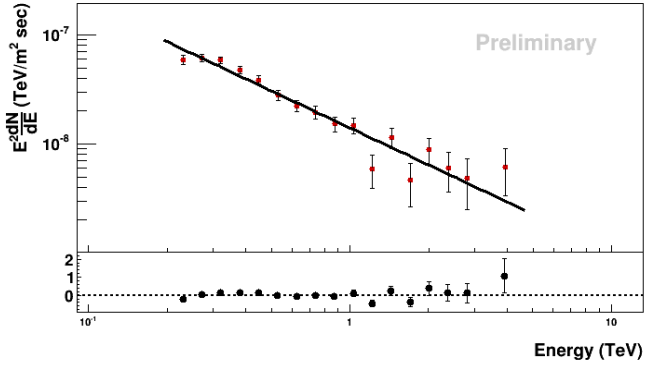}
  \caption{Time-averaged spectrum of 1ES1218+304.  Fit with a power law function between 220 GeV and 4 TeV.}
  \label{pic:1ES1218Spectrum}
 \end{figure}

 The spectrum of 1ES1218+304, shown in figure \ref{pic:1ES1218Spectrum}, was reconstructed between 220 GeV and 4 TeV.  A power law fit gives $F_{0}$ = 1.40 $\pm$ 0.10$_{stat}$ $\pm$ 0.30$_{sys}$ $\times$ 10$^{-8}$ photons per square meter per second per TeV and $\Gamma$ = 3.13 $\pm$ 0.05$_{sys}$ $\pm$ 0.20$_{sys}$, for a $\chi^{2}$ of 26.9 with 15 degrees of freedom.  Compared to the Crab Nebula spectrum derived from the same analysis chain, this corresponds to a flux of 8\% of the Crab flux above 200 GeV.  The brightness of this source allows for a high-statistics measurement of the spectrum.

This result agrees well with \cite{bib:1218VERITAS}, which gives a spectral index of $3.08 \pm 0.34_{stat}$, and \cite{bib:1218Asif}, which gives a spectral index of $\Gamma = 3.07 \pm 0.09_{stat}$.
 
\subsection{Analysis of 1ES0229+200}

1ES0229+200 is a relatively weaker source with a nonetheless hard spectrum.  The blazar is located at a redshift of $z=0.14$ \cite{bib:Woo2005}.  Observations have been taken from October 2009 (after the telescope 1 move) through January 2013.  54 hours out of 84 available hours of data remain after run selection and quality cuts. A total of 489 excess gamma-ray events were detected, with a cumulative statistical significance of 11.7 standard deviations above the background (\cite{bib:1ES0229ICRC}).  The spectrum was reconstructed between 280 GeV and 7.5 TeV, and fit with a power law function.  The resulting fit parameters are $F_{0}$ = 5.54 $\pm$ 0.56$_{stat}$ $\pm$ 1.10$_{sys}$ $\times$ 10$^{-9}$ photons per square meter per second per TeV and $\Gamma$ = 2.59 $\pm$ 0.12$_{stat}$ $\pm$ 0.20$_{sys}$, for a $\chi^{2}$ of 5.8 with 7 degrees of freedom.  This corresponds to 2\% of the Crab Nebula flux.

  \begin{figure}[t]
  \centering
  \includegraphics[width=0.5\textwidth]{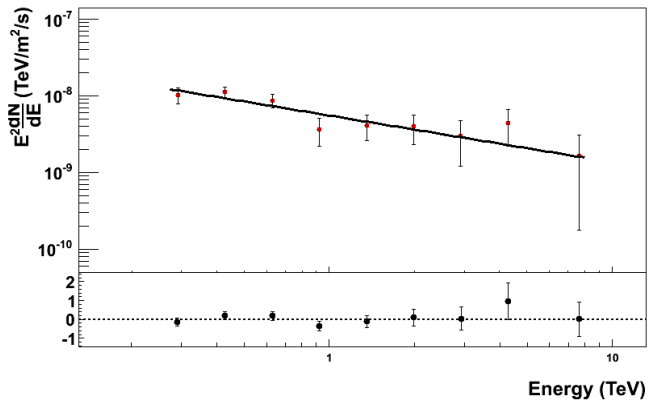}
  \caption{Time-averaged spectrum of 1ES0229+200.}
  \label{pic:1ES0229Spectrum}
 \end{figure}

Figure \ref{pic:1ES0229Spectrum} shows the spectrum of 1ES0229+200. 

\subsection{Analysis of 1ES0414+009}

The blazar 1ES0414+009 is located at a redshift of $z=0.290$ (\cite{bib:1ES0414} and has been observed from January 2008 until February 2011 for 84 hours.  A total of 56.2 hours of data remain after run selection and quality cuts.  A total of 822 excess gamma-ray events were detected, corresponding to a statistical significance of 6.4 standard deviations above the background.  The spectrum was reconstructed between 230 GeV and 850 GeV, and fit with the power law function, normalized at 300 GeV, such that $F(E) = F_{0}(\frac{E}{300 GeV})^{-\Gamma}$.  The resulting fit parameters are $F_{0}$ = 1.60 $\pm$ 0.30$_{stat}$ $\pm$ 0.80$_{sys}$ $\times$ 10$^{-7}$ photons per square meter per second per TeV and $\Gamma$ = 3.40 $\pm$ 0.50$_{stat}$ $\pm$ 0.30$_{sys}$, for a $\chi^{2}$ of 1.13 with 2 degrees of freedom.  This corresponds to a flux of 2\% of the Crab Nebula.

\begin{figure}[b]
  \centering
  \includegraphics[width=0.5\textwidth]{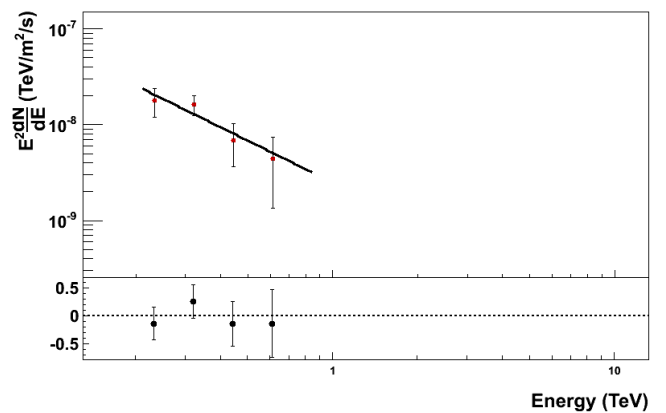}
  \caption{Time-averaged spectrum of 1ES0414+009.}
  \label{pic:1ES0414Spectrum}
 \end{figure}

Figure \ref{pic:1ES0414Spectrum} shows the spectrum of 1ES0414+009.

\section{Conclusions}

In this paper first results are shown from an ongoing VERITAS long term observation program targeting hard spectrum blazars.  These data cover several observing seasons and also include observations with the recently upgraded VERITAS camera trigger and the new high quantum efficiency photomultipliers  (see \cite{bib:ZitzerICRC} and \cite{bib:KiedaICRC}).   The energy spectra are unusually hard (2.60 - 3.45)  given their relatively large redshifts ranging of 0.14 - 0.28.  All of these spectra are well fit by power laws.  The blazars RGB J071+591 and H1426+428 have also been observed for long time periods (90 hours and 40 hours respectively). Future work will entail the analysis of deep observations on additional blazar sources, as well as a search for spectral breaks as a function of blazar redshift.

\begin{table}[h]
\begin{center}
\begin{tabular}{|p{2cm}|p{1.3cm}|p{1.7cm}|p{1.65cm}|}
\hline Source & Redshift & Detection Significance & Spectral Index \\ \hline
Crab Nebula   & N/A  & 95$\sigma$  & 2.50 $\pm$0.02\\ \hline
1ES1218+304   & 0.182  & 40$\sigma$  & 3.13 $\pm$0.05\\ \hline
1ES0229+200   & 0.140  & 12$\sigma$ & 2.59$\pm$0.12\\ \hline
1ES0414+009   & 0.287  & 6$\sigma$ & 3.40$\pm$0.50\\ \hline
\end{tabular}
\caption{Summary of Analysis.}
\label{tab:SummaryOfAnalysis}
\end{center}
\end{table}

\vspace*{0.5cm}
\footnotesize{{\bf Acknowledgments: }{This research is supported by grants from the U.S. Department of Energy Office of Science, the U.S. National Science Foundation and the Smithsonian Institution, by NSERC in Canada, by Science Foundation Ireland (SFI 10/RFP/AST2748) and by STFC in the U.K.  F.K. acknowledges support from the NASA Fermi Guest Investigator Grant NNX11A038. We acknowledge the excellent work of the technical support staff at the Fred Lawrence Whipple Observatory and at the collaborating institutions in the construction and operation of the instrument.}}


\begin{thebibliography}{}
\bibitem{bib:GouldSchreder} R.J. Gould and G. Schr\'{e}der, 1967. Phys. Rev. 155:5.

\bibitem{bib:KrennrichOrrEBLPaper} E. Dwek and F. Krennrich, 2012.  Astropar. Phys. 43:112-133.

\bibitem{bib:MayerEBLPaper} Meyer et al., 2012. A\&A, 542, A59.

\bibitem{bib:HessEBL} Abramowski et al., 2013. A\&A, 550, A4.

\bibitem{bib:AckermannEBLPaper} Ackermann et al., 2012. Science, 338.

\bibitem{bib:OrrEtAl} M. Orr, et al., 2011. ApJ. 733:77.

\bibitem{bib:HillasParameters} Hillas, M., 1985, Proc. 19th Int. Cosmic Ray Conf. (La Jolla, USA), 3, 445.

\bibitem{bib:WobbleAnalysis} Berge, D., Funk, S., \& Hinton, J., 2007, A\&A, 466, 1219.

\bibitem{bib:1218Asif} Acciari, V.A., et al., 2010. ApJ. 709 L 163.

\bibitem{bib:1218VERITAS} Acciari, V.A., et al., 2009. ApJ. 695:1370-1375.

\bibitem{bib:Woo2005} Woo, J. et al., 2005. ApJ 631 762.

\bibitem{bib:1ES0229ICRC} Cerruti, M., 2013. These proceedings.

\bibitem{bib:1ES0414} E. Aliu, et al., 2012. ApJ. 755 118.

\bibitem{bib:ZitzerICRC} Zitzer, B., \textit{The VERITAS Upgraded Telescope-Level Trigger Systems: Technical Details and Performance Characterization.} 2013. These proceedings.

\bibitem{bib:KiedaICRC} Kieda, D., 2013, These proceedings.


\end{thebibliography}
\end{document}